\documentclass[aps,twocolumn,prl,10pt,superscriptaddress]{revtex4-1}
\usepackage{amsmath}
\usepackage{dcolumn}
\usepackage{epsfig}
\usepackage{graphicx}
\usepackage{latexsym}
\usepackage{amssymb}
\usepackage{color}
\usepackage{amsfonts}
\usepackage{bm}

\newcommand{\pco}{Pr$_2$CuO$_4$}
\newcommand{\pcco}{Pr$_{2-x}$Ce$_x$CuO$_4$}
\newcommand{\pcod}{Pr$_{2}$CuO$_{4\pm\delta}$}
\newcommand{\pccod}{Pr$_{2-x}$Ce$_x$CuO$_{4\pm\delta}$}

\begin{document}
\title{Quantum oscillations suggest hidden quantum phase transition in the cuprate superconductor \pcod{}}

\author{Nicholas P. Breznay*}
\affiliation{Materials Science Division, Lawrence Berkeley National Laboratory, Berkeley CA 94720, USA and\\Department of Physics, University of California, Berkeley, Berkeley CA 94720, USA}
\altaffiliation{Contact for correspondence, nbreznay@berkeley.edu or analytis@berkeley.edu}
\author{Ross D. McDonald}
\affiliation{National High Magnetic Field Laboratory, Los Alamos National Laboratory, Los Alamos, New Mexico 87545, USA}
\author{Yoshiharu Krockenberger}
\affiliation{NTT Basic Research Laboratories, NTT Corporation, 3-1 Morinosato-Wakamiya, Atsugi, Kanagawa 243-0198, Japan}
\author{K. A. Modic}
\affiliation{Department of Physics, University of Texas, Austin TX 78712, USA}
\affiliation{National High Magnetic Field Laboratory, Los Alamos National Laboratory, Los Alamos, New Mexico 87545, USA}
\author{Zengwei Zhu}
\affiliation{National High Magnetic Field Laboratory, Los Alamos National Laboratory, Los Alamos, New Mexico 87545, USA}
\author{Ian M. Hayes}
\affiliation{Materials Science Division, Lawrence Berkeley National Laboratory, Berkeley CA 94720, USA and\\Department of Physics, University of California, Berkeley, Berkeley CA 94720, USA}
\author{Nityan L. Nair}
\affiliation{Materials Science Division, Lawrence Berkeley National Laboratory, Berkeley CA 94720, USA and\\Department of Physics, University of California, Berkeley, Berkeley CA 94720, USA}
\author{Toni Helm}
\affiliation{Materials Science Division, Lawrence Berkeley National Laboratory, Berkeley CA 94720, USA and\\Department of Physics, University of California, Berkeley, Berkeley CA 94720, USA}
\author{Hiroshi Irie}
\affiliation{NTT Basic Research Laboratories, NTT Corporation, 3-1 Morinosato-Wakamiya, Atsugi, Kanagawa 243-0198, Japan}
\author{Hideki Yamamoto}
\affiliation{NTT Basic Research Laboratories, NTT Corporation, 3-1 Morinosato-Wakamiya, Atsugi, Kanagawa 243-0198, Japan}
\author{James G. Analytis*}
\affiliation{Materials Science Division, Lawrence Berkeley National Laboratory, Berkeley CA 94720, USA and\\Department of Physics, University of California, Berkeley, Berkeley CA 94720, USA}

\date{\today}

\begin{abstract}
For both electron- and hole-doped cuprates, superconductivity appears in the vicinity of suppressed broken symmetry order, suggesting that quantum criticality plays a vital role in the physics of these systems. A confounding factor in identifying the role of quantum criticality in the electron-doped systems is the competing influence of chemical doping and oxygen stoichiometry. Using high quality thin films of \pcod, we tune superconductivity and uncover the influence of quantum criticality without Ce substitution.  We observe magnetic quantum oscillations that are consistent with the presence of small hole-like Fermi surface pockets, and a large mass enhancement near the suppression of superconductivity. Tuning these materials using only oxygen stoichiometry allows the observation of quantum oscillations and provides a new axis with which to explore the physics underlying the electron-doped side of the cuprate phase diagram.
\end{abstract}

\maketitle

Identifying the nature of ordered states that are proximal to superconductivity is central to our understanding of the cuprates. The properties of these states set important limits on their underlying microscopic description, and can reveal the presence of a quantum critical point and associated changes in symmetry and dominant interactions. Describing the physics common to both electron- and hole-doped cuprates is complicated in large part because of their different structures. Hole-doped systems (e.g., La$_{2-x}$Sr$_{x}$CuO$_{4}$ and YBa$_{2}$Cu$_{3}$O$_{7}$) have octahedrally or pyramidally coordinated Cu ions, while electron-doped systems such as Pr$_{2-x}$Ce$_{x}$CuO$_{4}$ (PCCO) have square planar coordinated Cu ions. Moreover, it is well known that electron-doped cuprates, unlike their hole-doped counterparts, do not exhibit superconductivity as grown, but need to be annealed in an oxygen deficient environment~\cite{tokura1989, takagi_superconductivity_1989}.

The effect of annealing and its relationship to chemical doping remains controversial, and has led to spirited debate over the electron-doped phase diagram~\cite{motoyama_spin_2007, li_impact_2008}. In particular, the questions of whether antiferromagnetism (AFM) co-exists with superconductivity or is phase separated by oxygen inhomogeneity, and whether a quantum critical point (QCP) exists near optimal $T_c$ ($x\sim 0.16$) or at much lower (Ce) doping ($x\leq 0.13$)~\cite{dagan2004, yu2007, motoyama_spin_2007, li_impact_2008, butch2012}, remain unresolved. A central difficulty in understanding simultaneously doped and annealed cuprates is the role of disorder, which is known to affect both magnetism and superconductivity~\cite{higgins2006, motoyama_spin_2007}. Magnetic quantum oscillations are exponentially sensitive to the electronic mean free path of a material~\cite{schoenberg}, and so cuprates that show quantum oscillations have facilitated rapid progress in our understanding of their physics~\cite{doiron2007, leboeuf2007,helm2009, helm2010, ramshaw2010}. In particular they have allowed the identification of the presence of a QCP in hole and electron-doped systems, indicated  by a strong enhancement in the effective mass as a function of doping~\cite{helm2015, ramshaw2015}. In this work, we utilize recent advances in thin film synthesis of \pco\, to make ultra-clean samples that are metallic and superconducting without addition of Ce \cite{higgins2006, krockenberger2013}, allowing us to observe quantum oscillations for the first time. This allows us to investigate the physics of electron-doped cuprates without the challenge of disentangling the effects of Ce substitution and annealing.

\begin{figure}[ht]
	\includegraphics[width=1.0\columnwidth]{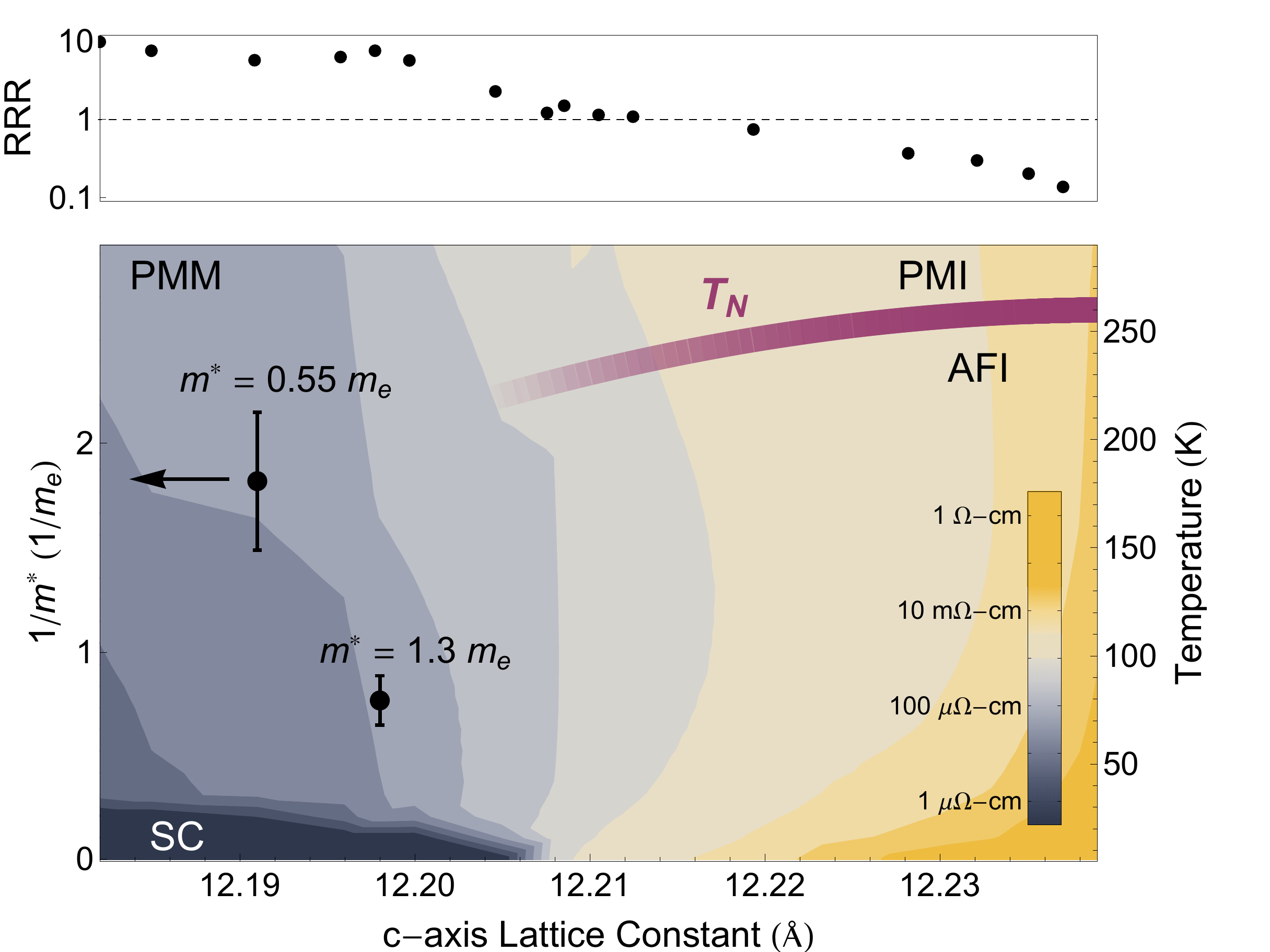}
	\caption{Transport in \pcod{} films. Contour plot of the in-plane resistivity $\rho_{xx}$ as a function of c-axis lattice parameter and temperature (right scale), measured on 18 different \pcod{} samples. Also shown (left scale) are the inverse quasiparticle effective masses determined from quantum oscillation measurements on two samples, discussed in the text, as well as a schematic boundary (thick line) between the antiferromagnetic insulating phase (AFI) as determined from bulk magnetic measurements of the Neel ordering temperature $T_N$~\cite{matsuda1990,*cox1989,*allenspach1989}, and the paramagnetic metallic (PMM), insulating (PMI), and superconducting (SC) phases. Top panel: residual resistivity ratio RRR, defined as $\rho(275 K)/\rho(30 K)$.}
	\label{f:phasediag}
\end{figure}

Our observation of quantum oscillations in high quality molecular beam epitaxy (MBE) grown films of \pcod\,serves as a direct probe of the Fermi surface  and quasiparticle interactions in this cuprate superconductor. We reveal a correlation between the electronic effective mass and the structural $c$-axis lattice constant, which can be tuned through oxygen annealing. Our main findings are summarized in Fig.~\ref{f:phasediag},
which shows the crossover from insulating to metallic temperature dependent in-plane resistivity $\rho_{xx}$ as a function of the interlayer lattice parameter $c$, as well as the inverse quasiparticle effective mass of two samples for which we have observed quantum oscillations and evidence for Fermi surface reconstruction. Also shown is the antiferromagnetic ordering temperature $T_N$ observed in bulk samples~\cite{matsuda1990,*cox1989,*allenspach1989}, and the residual resistivity ratio (RRR $\equiv \rho(275 K) / \rho(30 K)$) which crosses from metallic (RRR $\gtrsim 1$) to insulating (RRR $\ll 1$) behavior at $c=12.205$\,\AA{}.

\begin{figure*}[t]
    \includegraphics[width=2.0\columnwidth]{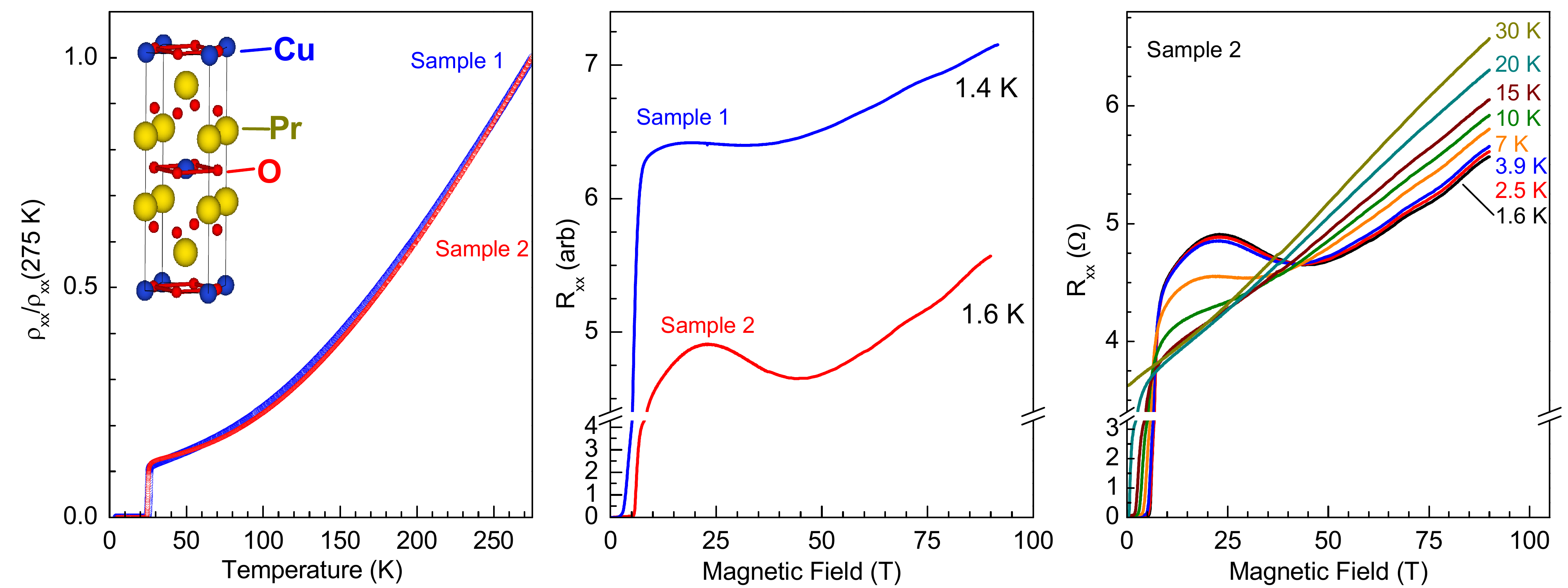}
    \caption{Resistivity and magnetotransport data. First panel: resistivity $\rho_{xx}$ versus temperature for two PCO samples. Inset: tetragonal unit cell of \pco{}. Second panel: magnetoresistance measured to 90 T for both samples near 1.5 K; note the break in axis scale. Third panel: suppression of both quantum oscillations, and a region of increased resistance at fields between 5 and 40 T, with temperature.}
\label{f:rvstmr}
\end{figure*}

Quantum oscillations are the unequivocal signature of a material's Fermi surface (FS), which in turn is the defining characteristic of a metal. Both thermodynamic and transport properties are acutely sensitive to the density of electronic states at this energy. In metals, the electronic states are quantized in the presence of a magnetic field $B$. For sufficiently clean metals the density of states (DOS) becomes an oscillatory function of $B$ and quantities sensitive to the DOS such as conductivity exhibit quantum oscillations periodic in $B^{-1}$. Quantum oscillations are often difficult to resolve in complex materials (such as transition metal oxides) due to disorder effects, which broaden the Landau levels until they are indistinguishable in laboratory scale magnetic fields. The discovery of quantum oscillations in hole-doped YBa$_2$Cu$_3$O$_{6.5}$\cite{doiron2007, leboeuf2007} opened the search for Fermi surface spectroscopy throughout the phase diagram of the cuprates. To date a limited range of hole dopings have been explored in YBa$_2$Cu$_3$O$_{6.5}$\cite{doiron2007, ramshaw2010}, YBa$_2$Cu$_4$O$_8$\cite{yelland2008}, and in HgBa$_2$CuO$_{4+\delta}$,\cite{barisic2013} and Tl$_2$Ba$_2$CuO$_{6+d}$ \cite{vignolle2008}. Both large and small Fermi surface pockets have been revealed near optimal $T_c$ in electron-doped Nd$_{2-x}$Ce$_{x}$CuO$_4$ (NCCO)\cite{helm2009}, and for both electron\cite{helm2015} and hole-doped\cite{ramshaw2015} cuprates the change of the Fermi surface with doping is accompanied by an evolution of the effective mass, and interpreted as arising from correlations near critical doping levels. For recent reviews see e.g. Refs.~\cite{vignolle2011,sebastian2012}.

\noindent \textbf{Film preparation and experimental techniques}

\noindent Films studied in this work are 100~nm thick, grown via molecular beam epitaxy on (001) SrTiO$_{3}$ substrates. Details of the synthesis, annealing, and extensive structural characterization of the films are discussed elsewhere\cite{krockenberger2013}. Aside from a small difference in the post-growth annealing time, the films studied in this work are prepared identically, and the only structural difference is variation in the c-axis lattice parameter (see Table~\ref{t:params}). Films were patterned into Hall bar geometries for in-$ab$-plane ($\rho_{xx}$) and Hall ($\rho_{xy}$) resistivity measurements using four-point low-frequency techniques at temperatures to 0.3\,K with DC magnetic fields to 16\,T, and pulsed magnetic fields to above 90\,T applied perpendicular to the $ab$-plane. Several Hall bar devices were patterned and measured from each film and showed identical behavior; data shown below are from two devices measured to the highest fields available. We find no evidence of inhomogeniety in film materials characterization or our transport studies; additional discussion of film characterization data are presented in the Supplementary Information. Changes to the oxygen stoichiometry affect both the carrier concentration of the material and the local copper coordination from square planar to pyramidal (or octahedral). As a result, in cuprates with square-planar coordinated Cu, the Cu-O bond length is 10\,\% longer compared to octahedral coordination, contracting the crystallographic $c$-axis. The $c$-axis length is therefore correlated with, and hence can be used as a measure of, oxygen stoichiometry (see Fig.~\ref{f:phasediag}).

\begin{table*}[hbt]
\begin{center}
\caption{Sample characterization parameters. Data shown for two \pcod{} samples measured in this work, as well as (for reference) a Ce-doped film synthesized using similar conditions. Parameters include film annealing conditions, fractional Ce content $x$, interlayer lattice parameter $c$, low-temperature (30\,K) resistivity $\rho_{30K}$, residual resistivity ratio (RRR), superconducting transition temperature T$_c$, and $\ell_{tr}$ the transport mean free path. Results from the quantum oscillation analysis are separated at right: $\ell_D$ the mean free path extracted using the Dingle temperature T$_D$, quantum oscillation frequency $F$, and the quasiparticle effective mass m$^*$.}
\begin{ruledtabular}
\begin{tabular}{c l c c c c c c c | c c c c}
Sample	&	Composition &	$x$	& T$_{\text{red}}$& $c$			& $\rho_{30K}$			& RRR & T$_{c}$	& $\ell_{tr}$	&	$\ell_D$	&	T$_{D}$	& $F$	&	m$^*$		\\
				&			&(Ce content)	& (K) &	(\AA)		&	($\mu\Omega$\,cm) 	&			&	(K)			&	(nm)					&	(nm)			& (K)			&	(T)	&	(m$_e$)	\\
\hline
1		&	\pcod{} &-	&	525 &	12.196	&	61	& 9.6	& 25.7		&	17	&	3.4					& 230	 		& 307$\pm$10		& 0.55$\pm$0.1	\\
2 	&	\pcod{} &- 	&	575 &	12.201	&	28	& 9.2 &	25.0		&	23	&	4.5					& 76			& 310$\pm$10		& 1.3$\pm$0.2		\\
(Ref. \onlinecite{krockenberger2012})	&	\pccod{} &0.14	& -	&	12.148	&	21		& 8.5	&	24.6	&	21.1	&	-		&	-	& -			& -			\\
\end{tabular}
\end{ruledtabular}
\label{t:params}
\end{center}
\end{table*}

The prevailing picture of Ce and oxygen doping (see e.g. Ref.~\onlinecite{armitage2010} for a thorough review) is that Ce is an electron donor, and oxygen a hole donor. However, changing the oxygen stoichiometry $\delta$ is known to cause multiple effects including stabilization of antiferromagnetic order, addition of charge carriers and tuning of disorder\cite{higgins2006, yu2007}. Based on the known variation of lattice structure with oxygen content, our \pcod{} films are stoichiometric to within $\delta \leq 0.04$\cite{Krockenberger2015}. The observation of quantum oscillations in the high field magnetotransport studies of these films indicates that disorder is minimal.

\begin{figure}[th]
  \includegraphics[width=1.0\columnwidth]{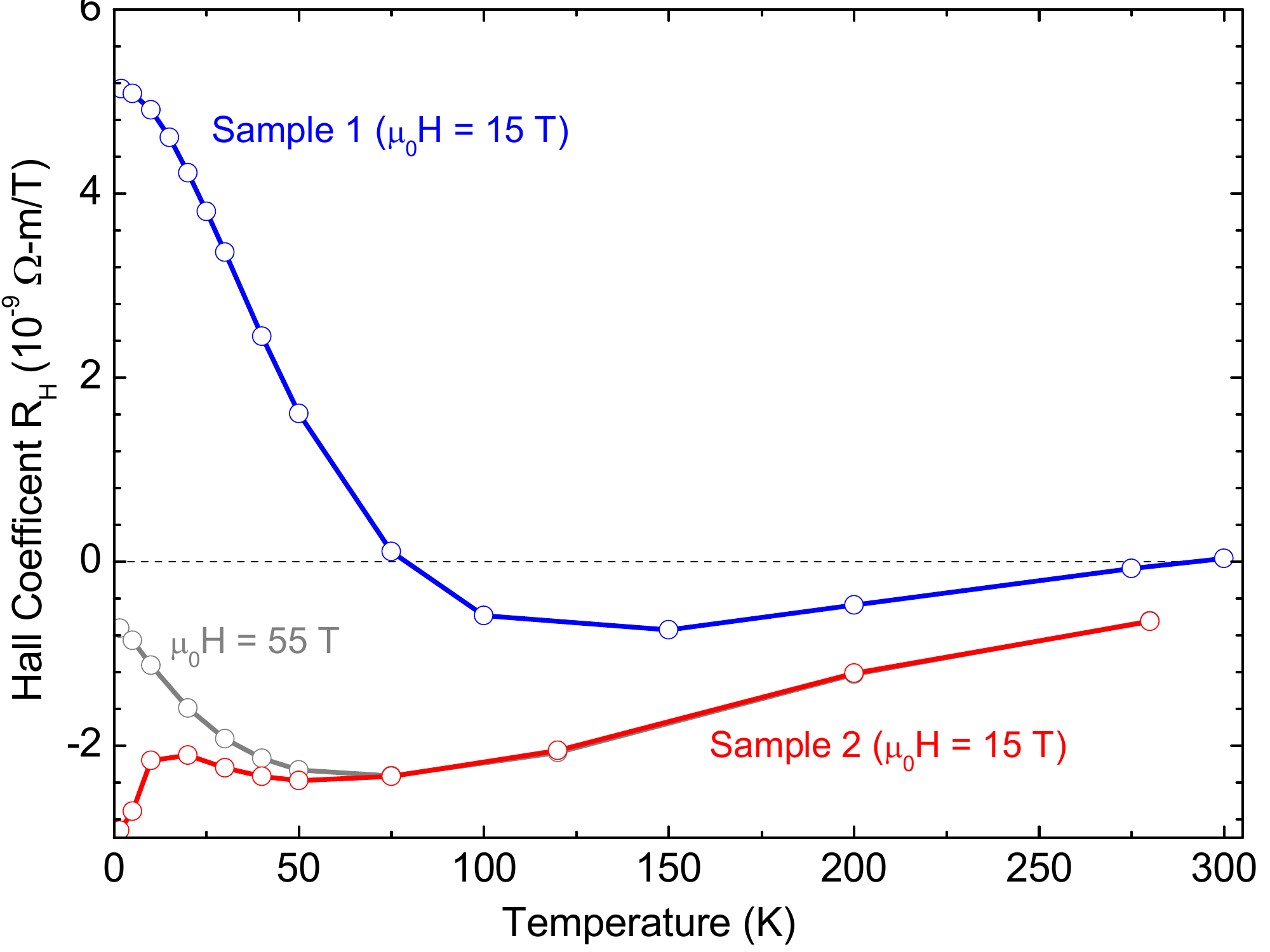}
  \caption{Hall coefficient $R_H$ versus temperature. Measurements were performed at magnetic fields of 15 T for both samples, and 55 T for sample 2.}
  \label{f:rh}
\end{figure}

\noindent \textbf{Magnetotransport results}

\noindent Figure~\ref{f:rvstmr} shows the temperature dependent resistivity of two \pco{} samples (1 and 2), measured in zero field;
the superconducting transition occurs at $T_c = 25$\,K for both films. The low-temperature residual resistivity (approximated by $\rho_{30 K}$, above the superconducting transition) is 30\,$\mu\Omega cm$ for sample 2. This is comparable to values observed in optimally doped \pcco{} films grown using various techniques\cite{charpentier2010,higgins2006,tafti2014}. An estimate of the carrier density from the low-temperature, low-field Hall coefficient (shown in Fig.~\ref{f:rh}) of $R_H = 5 \times 10^{-9}$\,$\Omega m/T$ for sample 1 gives $n = 1.0 \times 10^{22}$\,cm$^{-3}$. The change in sign of the Hall coefficient for sample 1 and strongly nonlinear Hall effect at low temperature (indicated by separation of the low-field and high-field R$_H$ in Fig.~\ref{f:rh} for sample 2) are inconsistent with a single, large-area FS. Instead, they indicate the presence of a broken-symmetry phase and reconstructed FS\cite{dagan2004,li2007}.

The low-temperature ($\sim$ 1.5 K) magnetoresistance for samples 1 and 2 are shown in Fig.~\ref{f:rvstmr}. Both samples show a sharp
transition to the normal state with a low-temperature upper critical field $B_{c2} = 6$\,T. Below T$_c$ the MR becomes negative for sample
1 (and strongly negative for sample 2) between 6.5\,T and $\sim$30\,T; such a ``hump'' in the MR has been linked to enhanced spin-dependent scattering~\cite{dagan2005}. After passing through a minimum, the magnetoresistance shows Shubnikov-de Haas quantum oscillations in the raw traces for both samples up to 90\,T\cite{twopoint}. At elevated temperatures the quantum oscillations are suppressed (visible for sample 2 in Fig.~\ref{f:rvstmr}) by thermal broadening within Landau levels; this suppression allows for an estimate of the effective mass.

\begin{figure}[ht]
  \includegraphics[width=1.0\columnwidth]{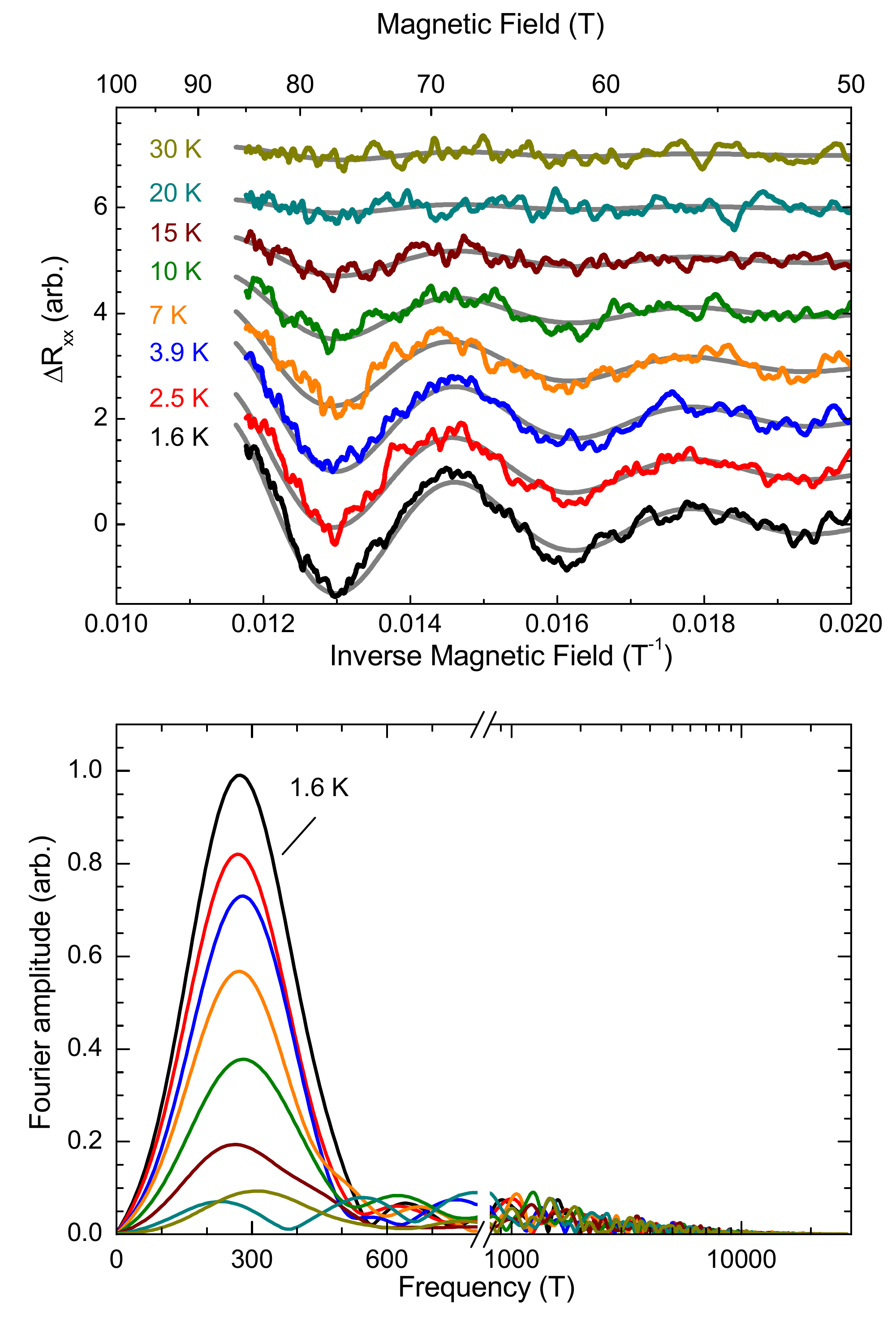}
	\caption{Magnetic quantum oscillations in \pcod{} sample 2. Top: resistivity versus field, along with fits to the Lifshitz-Kosevich formula (see text.) Bottom: fast Fourier transform spectra, showing a single peak near 300\,T (at left, linear frequency scale), and no evidence for high frequency oscillations (note break in axis, and log scale).}
	\label{f:qo}
\end{figure}

\noindent \textbf{Analysis of magnetic quantum oscillations}

\noindent To analyze the quantum oscillation spectrum, we subtract a smoothly varying low-order polynomial from the data shown in Fig.~\ref{f:rvstmr} to obtain the relative change in resistivity shown in Fig.~\ref{f:qo}. For such low frequency of oscillations, there is some sensitivity of subsequent analyses to the choice of background subtraction procedure; results described below have quoted uncertainties that reflect this sensitivity. Many background subtraction procedures were used to minimize systematic error in the determination of frequency and mass; details are described in the Supplementary Information. The first panel of Fig.~\ref{f:qo} shows the low-temperature magnetoresistance for sample 2 plotted versus inverse magnetic field, showing periodic oscillations with a frequency of 0.0032 T$^{-1}$. The Lifshitz-Kosevich description of the oscillating component of the magnetoresistance $\Delta R(B)$ of a quasi-2D Fermi surface is
\begin{equation}
\Delta R(B) \propto R_0 R_D R_T \cos\left( 2\pi F / B\right).
\label{eq:lk}
\end{equation}
where $R_0$ is an overall amplitude, $R_D = e^{- \pi / \omega_c \tau_D}$ is the Dingle factor, $R_T = \left( \frac{2\pi^2 k_B T/\hbar \omega_c}{\sinh\left( 2\pi^2 k_B T/\hbar \omega_c \right)} \right)$ is the thermal damping factor, $F$ is the quantum oscillation frequency, $\omega_c \equiv e B / \text{m}^*$ is the cyclotron frequency, and m$^*$ the is quasiparticle effective mass. Fits to Eq.~\ref{eq:lk} are shown in the first panel of Fig.~\ref{f:qo} and agree with the background-subtracted data with a single value of $F$ and the oscillation scattering time $\tau_D$.  The orbitally averaged mean free path $l_D=\tau_D\frac{\hbar}{\text{m}^*}\sqrt{\frac{F}{pi}} \approx 4$ nm is smaller than the transport mean free path $\ell_{tr} \approx 20$ nm for both samples, as expected since quantum oscillations are sensitive to both small and large angle scattering events. Sample mobilities $\mu$ are both $\sim 0.01$ T$^{-1}$ as calculated from both transport $\mu_{tr}=R_H/\rho_{xx}$ and quantum oscillation $\mu_D=\tau_D e/\text{m}^*$ measurements.

To precisely estimate $F$, we fit all of the quantum oscillatory data to a single temperature-independent frequency $F$, and a temperature-dependent amplitude. (Similar quantum oscillation data and analyses are shown for sample 1 in the Supplementary Information). The frequencies for the two samples are identical to within their uncertainties, $F_1 = 307$\,$\pm$\,10\,T and $F_2 = 310$\,$\pm$\,10\,T, as determined using fits to Eq.~\ref{eq:lk} and analyses of the fast Fourier transform (FFT) spectra. The FFTs for all temperatures are plotted in Fig.~\ref{f:qo} for sample 2, showing a single temperature-independent oscillation frequency near 300\,T.

The Onsager relation $F = \left( \frac{\hbar}{2 \pi e} \right) A_F$ relates the frequency $F$ to the extremal FS area $A_F$. We find $A_k = 2.8$ nm$^{-2}$; this is $\approx$ 1.1 \% of the Brillouin zone volume $A_{BZ}$ = $(2\pi/a)^2$ = 252 nm$^{-2}$ ($a$ is the in-plane lattice constant). This small BZ fraction is consistent with the reconstructed FS indicated by the Hall effect. No other frequencies are visible in either the raw signal or in the FFT (Fig.~\ref{f:qo}), including the vicinity of $\sim$10~kT (corresponding to $A_F \sim$0.4 $A_{BZ}$) that would be consistent with a large FS cylinder or magnetic breakdown as observed in NCCO \cite{helm2009}.

\begin{figure}[bh]
  \includegraphics[width=0.9\columnwidth]{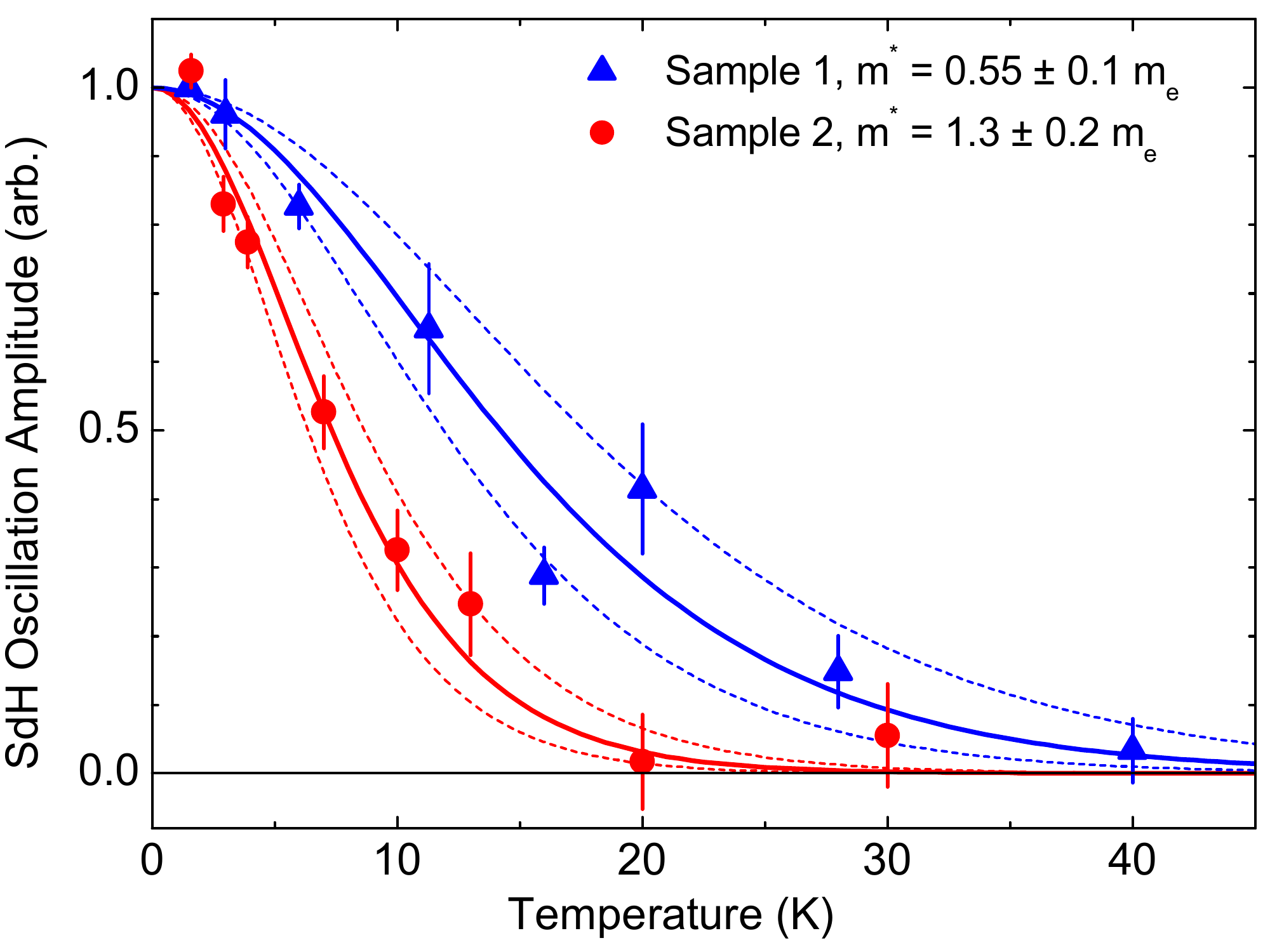}
	\caption{The quasiparticle effective mass. Quantum oscillation amplitude extracted from fast Fourier transform analyses versus temperature for two \pcod{} Hall bar devices; error bars indicate $\pm$1$\sigma$ standard deviation. Also shown are fits (continuous curves) to the Lifshitz-Kosevich temperature dependence for each sample (see text) along with broken lines indicating uncertainty in the best-fit value for m$^*$.}
	\label{f:mstar}
\end{figure}

From the temperature dependence of the quantum oscillation amplitudes ($R_T$ in Eq.~\ref{eq:lk}) we extract the quasiparticle effective mass m$^*$, which in general differs from the band mass due to the renormalization of many-body effects. The amplitude is plotted versus temperature in Fig.~\ref{f:mstar}, along with a single-parameter fit to the above expression yielding m$^* =  0.55 \pm 0.1$\,$m_e$ for sample 1, and m$^* =  1.3 \pm 0.2$\,$m_e$ for sample 2.

\noindent \textbf{Discussion and implications}

\noindent 

The effective mass changes by greater than a factor of two, with no discernible change in the FS cross-section (i.e. in the observed frequency $F$). It is unlikely that such a large change in effective mass could originate from changes in the band structure or band filling caused by a slight change in either doping or disorder driven by oxygen stoichiometry. A more plausible scenario is that this change in effective mass is the result of a change in the renormalizing many-body interactions. The strong variation in m$^*$ for the two samples studied here indicates that this material lies in close proximity to a quantum critical point on the boundary between ground states with competing symmetries. As shown in Fig.~\ref{f:phasediag}, we identify the c-axis lattice constant, similar to Ce content $x$ in \pcco{}, as a parameter identifying proximity to one or more broken-symmetry phases. Such proximity has been seen in the hole-doped cuprates, which evidence an electron-like pocket at low doping that evolves to a large hole-like pocket for overdoped materials (as observed via quantum oscillations in Tl$_2$Ba$_2$CuO$_{6+d}$ \cite{vignolle2008} and angle-dependent magnetotransport\cite{hussey2003}), with one or more quantum critical points in between\cite{ramshaw2015,helm2015}. Surface sensitive probes such as ARPES have observed analogous evolution with Ce doping in electron-doped materials such as NCCO~\cite{armitage2002,armitage2003,matsui2007}, while recent work has revealed the importance of oxygen reduction~\cite{song2012} in the evolution of apparent FS arcs near ``optimal'' doping. Our finding of a broken-symmetry reconstructed FS in \pcod{} suggests that such evolution may occur with oxygen reduction for a wide range of Ce content.

The frequency we observe is similar to that found in bulk samples of NCCO near optimal doping\cite{helm2009, helm2010, helm2015}, suggesting that the low-temperature ordered state in the films studied here is the same. Oscillations arising from a large FS cylinder as reported in ARPES~\cite{armitage2002,song2012}, or ascribed to magnetic breakdown in NCCO \cite{helm2015}, would appear with $F\sim$10\,kT but were not evident in either sample at temperatures down to 0.3\,K and fields to 90\,T. Our work demonstrates that MBE grown films are a new platform for studying the Fermi surface of cuprates and provide an unique method by which to cleanly tune these materials across a phase diagram without Ce doping. In particular, sensitivity to the effects of a nearby QCP in \pcod{} opens the exciting possibility of conclusively identifying not only the existence of a QCP, but of the pertinent tuning parameter that leads to high temperature superconductivity in these materials.

\begin{acknowledgments}
\textit{Acknowledgments-} We thank the scientific and support staff of the Los Alamos National High Magnetic Lab, Pulsed Field Facility
for their technical assistance and fruitful discussions throughout this project, and also acknowledge illuminating discussion with B.~
Ramshaw, N.~Harrison, R.~Greene, and M.~Naito. This work was supported by the Laboratory Directed Research and Development Program of Lawrence Berkeley National Laboratory under the US Department of Energy Contract No. DE-AC02-05CH11231. A portion of this work was performed at the National High Magnetic Field Laboratory, which is supported by the National Science Foundation Cooperative Agreement No. DMR-1157490, the State of Florida, and the U.S. Department of Energy. RDM and KAM acknowledge support from the DOE BES under ``Science of 100 tesla'' and ZZ acknowledges LANL directors funding under LDRD 20120772. We also acknowledge the support of the Gordon and Betty Moore Foundation.
\end{acknowledgments}


\begin{references}
\bibitem{twopoint}
Magnetoresistance data were measured in both 2-point and 4-point geometries and
  results were identical. All the high field data presented in this work come
  from 2-point measurements, and care was taken to guarantee that no lead
  resistance effects appear in all such measurements.
\end{references}
\bibliographystyle{unsrt}

\end{document}